\documentclass[a4paper,unpublished]{quantumarticle}
\pdfoutput=1

\usepackage[utf8]{inputenc}
\usepackage[T1]{fontenc}
\usepackage[english]{babel}
\usepackage{amsmath}
\usepackage{amssymb}
\usepackage{amsthm}
\usepackage{bm}
\usepackage{booktabs}
\usepackage{graphicx}
\usepackage{mathtools}
\usepackage{microtype}
\usepackage{xcolor}
\usepackage{tikz}
\usetikzlibrary{arrows.meta,fit,positioning}
\usepackage{listings}
\usepackage{xspace}
\usepackage[numbers,sort&compress]{natbib}
\usepackage[colorlinks=true,citecolor=quantumviolet,linkcolor=quantumviolet,urlcolor=quantumviolet]{hyperref}
\usepackage[nameinlink,capitalize]{cleveref}

\newcommand{\E}{\mathbb{E}}
\newcommand{\Prob}{\mathbb{P}}
\newcommand{\PL}{P_{\mathrm{L}}}
\newcommand{\FS}{FaultScope\xspace}
\newcommand{\Stim}{Stim\xspace}
\newcommand{\code}[1]{\texttt{#1}}

\begin{document}

\title{FaultScope: A software package for fast error attribution in quantum error correction}
\author{Yuguo Shao}
\thanks{These authors contributed equally to this work.}
\affiliation{Yau Mathematical Sciences Center, Tsinghua University, Beijing 100084, China}
\affiliation{Beijing Institute of Mathematical Sciences and Applications, Beijing 100407, China}
\author{Zhenyu Chen}
\thanks{These authors contributed equally to this work.}
\affiliation{Department of Computer Science and Technology, Tsinghua University, Beijing 100084, China}
\author{Ningfeng Wang}
\affiliation{Yau Mathematical Sciences Center, Tsinghua University, Beijing 100084, China}
\author{Song Cheng}
\email{chengsong@bimsa.cn}
\affiliation{Beijing Institute of Mathematical Sciences and Applications, Beijing 100407, China}
\author{Zhaohui Wei}
\email{weizhaohui@gmail.com}
\affiliation{Yau Mathematical Sciences Center, Tsinghua University, Beijing 100084, China}
\affiliation{Beijing Institute of Mathematical Sciences and Applications, Beijing 100407, China}
\author{Zi-Wen Liu}
\email{zwliu0@tsinghua.edu.cn}
\affiliation{Yau Mathematical Sciences Center, Tsinghua University, Beijing 100084, China}
\author{Zhengwei Liu}
\email{liuzhengwei@bimsa.cn}
\affiliation{Yau Mathematical Sciences Center, Tsinghua University, Beijing 100084, China}
\affiliation{Beijing Institute of Mathematical Sciences and Applications, Beijing 100407, China}

\date{September 23, 2026}

\begin{abstract}
We introduce \FS, a classical simulator for quantum error correction circuits, implemented in Rust with a Python interface.
\FS estimates logical error rates and their derivatives with respect to component error rates from the same sampled and decoded shots, without modifying or differentiating through the decoder.
These derivatives enable error attribution to individual noisy components, helping identify noise hotspots for targeted noise reduction.
\FS supports forward simulation of error propagation through circuits and direct sampling of detector and logical observable outcomes from detector error models.
In our noiseless and noisy benchmarks, \FS has the lowest runtime among the tested simulators at every size.
Computing all derivatives adds only 0.33--0.97\% to the runtime of logical error rate estimation across five repetition code sizes with 7--2047 error locations.
\FS can be used to benchmark large-scale quantum error correction schemes and guide targeted noise reduction efficiently, thereby facilitating the design and optimization of fault-tolerant quantum computing architectures.
\end{abstract}

\maketitle

\section{Introduction}
\label{sec:introduction}

Classical simulation is a key tool for evaluating the performance of quantum error correction (QEC) schemes and decoders~\cite{terhal2015quantum}.
Stabilizer methods for Clifford circuits \cite{gottesman1997stabilizer,aaronson2004improved,anders2006fast} and direct sampling from detector error models (DEMs) \cite{gidney2021stim} support efficient generation of detector records and logical readouts across many shots.
The logical error rate estimated from these records after decoding measures overall failure probability, but does not reveal how individual noisy components contribute to logical failures~\cite{shao2025diagnosing,acharya2023suppressing,acharya2025belowthreshold}.
Error frequencies alone  are not enough: a frequent error may be corrected, whereas a rarer one may cause the decoder to choose the wrong logical correction.

Ref.~\cite{csim07} introduces an error attribution scheme that identifies where targeted noise reduction can most effectively improve QEC performance. The scheme estimates derivatives of the logical error rate with respect to the error rate of each noisy component. Each derivative measures the difference between the conditional probabilities of logical failure with and without an error at that component. We refer to these derivatives as sensitivities.
These estimates induce a hotspot map of the noisy components. When intervention resources are limited, the map guides targeted noise reduction on hotspot components to lower the logical error rate more effectively.
Circuit-level simulations of surface codes and high-rate lifted-product codes show that targeting these hotspots reduces the logical error rate substantially more than applying the same intervention to an equal number of randomly selected components.

To implement this scheme efficiently, we develop \FS, an open-source software package that estimates the logical error rate and component sensitivities from the same simulation records.
The simulation core is implemented in Rust and exposed through a Python interface.
It supports two simulation paths: forward simulation of Clifford circuits, and direct sampling from detector error models.
Both paths estimate all sensitivities from a single batch of sampled errors and decoded outcomes, without differentiating through or modifying the decoder.
\FS stores these records in a compact format and supports sensitivity summaries by qubit, gate, round, or operation, allowing noise hotspots to be examined at different levels of the circuit.
To assess the computational overhead of error attribution, we benchmark the complete sampling, decoding, and estimation workflow.
For repetition code circuits with 7--2047 noisy components and 100,000 shots per run, computing all sensitivities increases median total runtime by 0.33--0.97\% relative to logical error rate estimation without event recording.
On random Clifford circuits and noisy surface code memories, each using 1,000,000 shots per circuit and comparing six CPU simulators, \FS achieves the lowest observed median runtime at every circuit size with a completed comparison.
For random Clifford circuits with 1024 qubits and 64 layers, it achieves a median speedup of $1.70\times$ over \Stim across matched instances.
It is also the only tested simulator to complete the largest surface code case, at $d=100$, with circuit level Pauli noise at $p=10^{-3}$, under the benchmark resource limits. Our implementation and benchmarks demonstrate that error attribution can be integrated efficiently into QEC simulation, making it practical to identify noise hotspots and guide targeted improvements alongside routine performance evaluation.

\Cref{sec:attribution} defines the simulation model and estimated quantities.
\Cref{sec:implementation} describes the execution paths and computational cost.
\Cref{sec:benchmark} presents performance benchmarks, and \cref{sec:discussion} discusses the results and limitations.

\section{Preliminaries}
\label{sec:attribution}

\subsection{Error model and logical failure rate}

We consider QEC memory experiments specified by a Clifford syndrome extraction circuit, a stochastic Pauli error model, and a decoder $\mathcal D$~\cite{fowler2012surface}.
In each run, a logical qubit is initialized in a chosen logical $X$- or $Z$-basis state and undergoes one or more rounds of syndrome extraction.
Final measurements of the data qubits provide the logical readout in the same basis.
The error model assigns $M$ error events to circuit operations and locations, with independent Bernoulli occurrences parameterized by the probabilities $\bm p=(p_1,\ldots,p_M)$.
For shot $i$, let $\bm d_i$ be the syndrome record constructed from the measurement outcomes and let $o_i\in\{0,1\}$ indicate whether the final logical readout differs from its noiseless value.
The decoder uses $\bm d_i$ to infer a correction $c_{\mathcal D}(\bm d_i)\in\{0,1\}$, which specifies whether the logical readout should be flipped.
A logical failure occurs if the corrected readout still differs from the noiseless value, giving the failure indicator
\begin{equation}
  L_i=\bm 1\!\left[o_i\oplus c_{\mathcal D}(\bm d_i)=1\right].
  \label{eq:loss}
\end{equation}
Here $\oplus$ is addition modulo two, and $\bm 1[\cdot]$ is the indicator function.
For a fixed circuit and decoder, the logical error rate is the probability of this failure over repeated runs:
\begin{equation}
  \PL(\bm p;\mathcal D)
  =\Prob_{\bm p}(L_i=1)=\E_{\bm p}[L_i].
  \label{eq:logical-error-rate}
\end{equation}

\subsection{Sensitivities and hotspots}

The sensitivity $S_j$ is the derivative of the logical error rate with respect to the error rate $p_j$~\cite{csim07}:
\begin{equation}
  S_j=\frac{\partial\PL(\bm p;\mathcal D)}{\partial p_j}.
  \label{eq:sensitivity}
\end{equation}
The circuit, decoder, and all other error rates are held fixed when taking this derivative.

Let $F_{ij}=1$ if event $j$ occurs in shot $i$, and $F_{ij}=0$ otherwise.
Under the independent Bernoulli event model above, for $0<p_j<1$, the sensitivity can also be written as:
\begin{equation}
  S_j=\Prob(L_i=1\mid F_{ij}=1)-\Prob(L_i=1\mid F_{ij}=0).
  \label{eq:conditional-sensitivity}
\end{equation}
Thus $S_j$ compares the logical failure probabilities when event $j$ occurs and when it does not, averaged over all other sampled choices.

Mapping these sensitivities to their source components produces a noise hotspot map, with large positive sensitivities identifying errors whose occurrence substantially increases the probability of logical failure.
A physical component, such as a qubit or a two-qubit coupling, may be used many times in the circuit.
Let $G$ denote the set of error events associated with that component.
If all events in $G$ share the same error rate, the sensitivity of the physical component is
\begin{equation}
  S_G=\sum_{j\in G}S_j.
  \label{eq:component-sensitivity}
\end{equation}
For small reductions $\delta p_j\geq0$ in the error rates, the predicted decrease in logical error rate is $\sum_j S_j\delta p_j$.
These predictions, together with achievable noise reductions and intervention costs, guide the choice of components to target under a limited budget.

\subsection{Monte Carlo estimates and grouped outputs}

To estimate the logical error rate and its sensitivities, we simulate $N$ independent shots of the QEC memory experiment.
For each shot $i$, we record the error occurrences $F_{ij}$ and determine the logical failure indicator $L_i$ after decoding.
For $0<p_j<1$, these records give the estimates
\begin{align}
  \widehat P_{\mathrm L}&=\frac{1}{N}\sum_{i=1}^N L_i,
  \label{eq:logical-error-estimate}\\
  \widehat S_{j,b}
  &=\frac{1}{N}\sum_{i=1}^{N}(L_i-b)
  s_j(F_{ij}),
  \label{eq:mc-estimator}
\end{align}
where $b$ is a reference value, called a baseline, subtracted from $L_i$ to reduce the variance of the sensitivity estimate.
The factor $s_j(F_{ij})$, defined as
\begin{equation}
  s_j(F_{ij})=\frac{F_{ij}}{p_j}-\frac{1-F_{ij}}{1-p_j},
  \label{eq:likelihood-score}
\end{equation}
is called the likelihood score.
The decoder is used only to obtain the failure indicators $L_i$, so sensitivity estimation does not require differentiating through it.

The likelihood score has zero expectation, so subtracting a fixed baseline, or one estimated from independent data, leaves the estimator unbiased: $\E[\widehat S_{j,b}]=S_j$~\cite{mohamed2020montecarlo}.
For fixed $b$, the variance is
\begin{equation}
  \operatorname{Var}(\widehat S_{j,b})
  =\frac{\E[(L_i-b)^2s_j(F_{ij})^2]-S_j^2}{N}.
  \label{eq:sensitivity-variance}
\end{equation}
At a fixed shot count, a suitable baseline can reduce the variance of the sensitivity estimate.
For component $j$, the optimal constant baseline that minimizes this variance~\cite{greensmith2004variance} is
\begin{equation}
  b_j^{\star}
  =\frac{\E[L_i s_j(F_{ij})^2]}{\E[s_j(F_{ij})^2]}.
  \label{eq:optimal-baseline}
\end{equation}

To estimate the sensitivity of a physical component associated with event set $G$, we combine the event estimates according to \cref{eq:component-sensitivity}:
\begin{equation}
  \widehat S_{G,b}=\sum_{j\in G}\widehat S_{j,b}.
  \label{eq:estimated-component-sensitivity}
\end{equation}
The event estimates can also be organized by qubit, gate, or syndrome extraction round to construct noise hotspot maps at different levels of the circuit.
For experiments with multiple logical observables, the loss can be defined separately for each observable or as a combined failure indicator.
The same estimation procedure applies to other user defined losses, provided that their definitions remain unchanged as the error rates vary.

\section{Software design and implementation}
\label{sec:implementation}

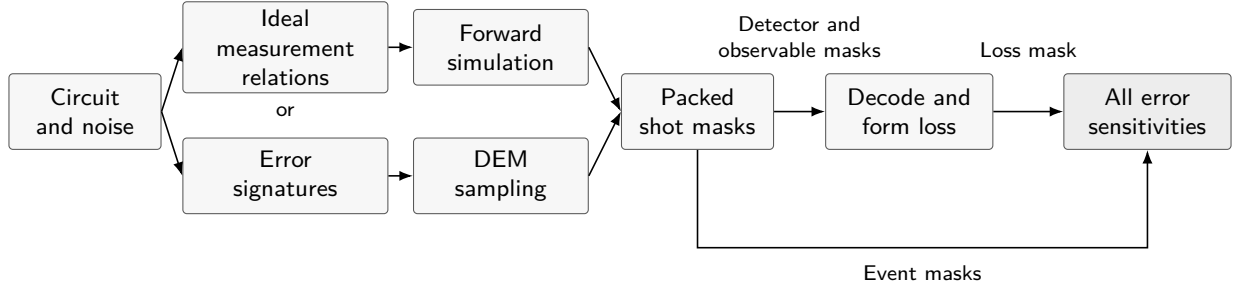
\begin{figure*}[t]
  \centering
  \begin{tikzpicture}[
      font=\small\sffamily,
      box/.style={draw=black!65,rounded corners=1.5pt,align=center,
                  minimum height=10mm,text width=25mm,inner sep=3pt,
                  fill=black!3},
      arrow/.style={-{Latex[length=1.8mm]},line width=0.6pt},
      flowlabel/.style={font=\footnotesize\sffamily,align=center}
    ]
    \node[box,text width=18mm] (input) at (0,0) {Circuit\\and noise};
    \node[box,text width=25mm] (ideal) at (2.65,0.85)
      {Ideal\\measurement relations};
    \node[box,text width=25mm] (signatures) at (2.65,-0.85)
      {Error\\signatures};
    \node[flowlabel] at (2.65,0) {or};
    \node[box,text width=21mm] (forward) at (5.5,0.85)
      {Forward\\simulation};
    \node[box,text width=21mm] (dem) at (5.5,-0.85) {DEM\\sampling};
    \node[box,text width=18mm] (records) at (8.1,0) {Packed\\shot masks};
    \node[box,text width=20mm] (decoder) at (10.9,0) {Decode and\\form loss};
    \node[box,text width=20mm,fill=black!7] (score) at (14.05,0)
      {All error\\sensitivities};

    \draw[arrow] (input.east) -- (ideal.west);
    \draw[arrow] (input.east) -- (signatures.west);
    \draw[arrow] (ideal.east) -- (forward.west);
    \draw[arrow] (signatures.east) -- (dem.west);
    \draw[arrow] (forward.east) -- (records.west);
    \draw[arrow] (dem.east) -- (records.west);
    \draw[arrow] (records.east) --
      node[flowlabel,above=6mm] {Detector and\\observable masks}
      (decoder.west);
    \draw[arrow] (decoder.east) --
      node[flowlabel,above=6mm] {Loss mask} (score.west);
    \draw[arrow] (records.south) -- (8.1,-1.8) --
      node[flowlabel,below=1mm] {Event masks} (14.05,-1.8) -- (score.south);
  \end{tikzpicture}
  \caption{Architecture of \FS.
  Compilation precomputes ideal measurement relations for forward simulation or
  error signatures for DEM sampling; either representation is reused across
  shot batches. During batch execution, event masks bypass decoding and are
  combined with the loss mask from the same batch to estimate all sensitivities.}
  \label{fig:architecture}
\end{figure*}

The estimators in \cref{sec:attribution} use the error occurrences and logical failure indicators recorded for each shot.
To generate and process these records efficiently, \FS compiles a circuit into either a forward stabilizer program or a DEM sampling plan (\cref{fig:architecture}).
Both paths use the same data format and share the decoding and sensitivity estimation stages.

In both execution paths, the binary outcomes of each detector over $N$ shots form an $N$-bit vector.
Each logical observable and each tracked error event has its own vector of the same form.
We call each such vector a \emph{shot mask}.
The \emph{event mask} for event $j$ is $(F_{1j},\ldots,F_{Nj})$, with bit $i$ indicating whether that event occurred in shot $i$.
The observable mask records whether the logical readout in each shot differs from its noiseless value.
Each mask is stored in $\lceil N/64\rceil$ 64-bit words.

\subsection{Forward stabilizer simulation}

The forward path combines stabilizer simulation of the ideal Clifford circuit \cite{gottesman1997stabilizer,aaronson2004improved} with Pauli frame propagation of sampled Pauli errors \cite{gidney2021stim}.
A \emph{Pauli frame} records the accumulated Pauli error on each qubit; Clifford gates propagate it by mapping Pauli operators to Pauli operators under conjugation~\cite{dehaene2003clifford}.
\FS first compiles the ideal measurement relations, then propagates the sampled Pauli frame through the circuit, and finally combines the ideal outcomes with the flips caused by the frame to generate measurement records.

During compilation, \FS analyzes the noiseless circuit using a stabilizer representation.
It derives an expression for each ideal measurement outcome in terms of fixed bits and independent random bits that will be sampled at runtime.
Consider a single shot, with the shot index omitted.
For measurement $k$ in the circuit, $m_k^{(0)}\in\{0,1\}$ denotes the ideal outcome in the absence of stochastic errors, and $m_k\in\{0,1\}$ denotes the corresponding noisy outcome in the presence of sampled Pauli errors.
The ideal outcome is expressed as an XOR of a constant bit and selected random bits, and the noisy outcome is obtained by applying the flip caused by the sampled Pauli frame:
\begin{equation}
  m_k^{(0)}=c_k\oplus\bigoplus_{r\in A_k}\xi_r,
  \qquad
  m_k=m_k^{(0)}\oplus\phi_k(\bm{x},\bm{z}),
  \label{eq:reference-frame-measurement}
\end{equation}
Here $c_k\in\{0,1\}$ is a constant bit and $A_k$ contains the indices of the random bits contributing to measurement $k$; both are determined during compilation.
The independent bits $\xi_r$ represent the intrinsic randomness of quantum measurements in the ideal circuit and are sampled as 0 or 1 with equal probability for each shot.

For each shot, Pauli errors are sampled from the noise model and propagated from their circuit locations through subsequent Clifford gates.
The binary vectors $\bm{x}$ and $\bm{z}$ record the resulting Pauli frame immediately before measurement $k$, with one entry per qubit for its $X$ and $Z$ components, respectively.
The flip indicator $\phi_k(\bm{x},\bm{z})\in\{0,1\}$ equals 1 if this frame anticommutes with the measured Pauli operator, thereby flipping the ideal outcome, and 0 otherwise.

For a single shot, $x_q,z_q\in\{0,1\}$ encode the Pauli error on qubit $q$ as $X_q^{x_q}Z_q^{z_q}$, up to an overall phase.
At run time, the $x_q$ values from all shots are packed into one mask, and the $z_q$ values into another.
The same notation $x_q,z_q$ is used for these masks in the updates below.
Random bits and measurement outcomes are packed in the same way.

An $H$ gate interchanges the $X$ and $Z$ components and therefore swaps $x_q$ and $z_q$.
Since $S$ maps $X$ to $Y$ and leaves $Z$ unchanged, its mask update is $z_q\leftarrow z_q\oplus x_q$.
For CNOT with control $c$ and target $t$, the $X$ component propagates from control to target and the $Z$ component from target to control, giving $x_t\leftarrow x_t\oplus x_c$ and $z_c\leftarrow z_c\oplus z_t$.
Each 64-bit XOR processes up to 64 shots at once.
For $H$ and qubit SWAP, the required mask exchanges can be implemented by swapping buffer references, avoiding copies of the stored bits.

\begin{table*}[!t]
  \centering
  \small
  \caption{Performance optimizations in \FS, grouped by the cost term in
  \cref{eq:attribution-cost} that they reduce. $M$: number of tracked error
  events; $N$: number of shots; $B=\lceil N/64\rceil$: words per shot mask.}
  \label{tab:performance-mechanisms}
  \begin{tabular}{p{0.19\textwidth}p{0.49\textwidth}p{0.25\textwidth}}
    \toprule
    \raggedright Optimization
      & \raggedright What it does & \raggedright Scope and limits \tabularnewline
    \midrule
    \multicolumn{3}{l}{\emph{One time compilation (excluded from $T_{\mathrm{score}}$)}} \tabularnewline
    \addlinespace[2pt]
    \raggedright Symbolic ideal outcomes
      & \raggedright Each ideal measurement outcome is compiled once into the
        form of \cref{eq:reference-frame-measurement}. Noiseless Pauli gates
        are absorbed into the constant bits $c_k$.
      & \raggedright Forward path. \tabularnewline
    \raggedright Packed stabilizer tableau
      & \raggedright The tableau is stored column-wise in packed words, so each
        gate is a few word-wide XORs. The row form needed for measurements is
        rebuilt only when it is out of date. Keeping the destabilizers lets
        deterministic outcomes be read off without Gaussian elimination.
      & \raggedright Forward compilation. Single qubit measurements and resets
        use a further specialized path. \tabularnewline
    \raggedright Reuse of repeated structure
      & \raggedright A reset that immediately follows a measurement of the same
        qubits reuses the outcomes just computed. For a repeated loop body
        whose Pauli support returns to its initial value after one or two
        iterations, later iterations are generated from an affine template
        instead of being re-simulated.
      & \raggedright Forward compilation; applies only to exactly matching
        blocks or eligible repeats. \tabularnewline
    \raggedright Batched fault propagation
      & \raggedright To build $\Omega$ in \cref{eq:dem-matrix}, many elementary
        errors are propagated through the circuit together, one per bit.
        The resulting signatures replace circuit propagation in all later
        sampling batches.
      & \raggedright DEM path; requires independent events and supported
        circuit and noise semantics. \tabularnewline
    \midrule
    \multicolumn{3}{l}{\emph{Sampling, $T_{\mathrm{sample}}(N)$}} \tabularnewline
    \addlinespace[2pt]
    \raggedright Shot-parallel Pauli frames
      & \raggedright Frame bits of 64 shots share one word, so each Clifford
        gate costs $O(B)$ word XORs. $H$ and SWAP only exchange buffer
        pointers, at $O(1)$ cost.
      & \raggedright Forward path. \tabularnewline
    \raggedright Sparse event sampling
      & \raggedright Instead of drawing one random number per shot, geometrically
        distributed gaps jump directly to the shots in which event $j$ occurs,
        with expected $O(1+Np_j)$ work. Each occurrence flips only the
        detectors and observables in the event's signature.
      & \raggedright DEM path; buffer initialization and signature updates add
        fixed overhead. \tabularnewline
    \raggedright Output only what is used
      & \raggedright The DEM sampler writes directly in the decoder's input
        layout, and only requested measurement records are exported.
      & \raggedright Gain depends on the decoder interface. \tabularnewline
    \midrule
    \multicolumn{3}{l}{\emph{Attribution, $O(MB)$}} \tabularnewline
    \addlinespace[2pt]
    \raggedright One batch for all events
      & \raggedright All $M$ sensitivities are estimated from the same $N$
        decoded shots. Each needs only an AND with the loss mask and a
        population count over $B$ words.
      & \raggedright Storing all event masks takes $O(MB)$ words; they are not
        kept when attribution is disabled. \tabularnewline
    \bottomrule
  \end{tabular}
\end{table*}

\subsection{Detector error model sampling}

A DEM represents noise as independent events with probabilities $p_j$, each with a fixed \emph{signature} specifying which detectors and logical observables it flips \cite{gidney2021stim}.
\FS compiles these signatures once and uses them to generate detector and observable records by sampling event occurrences.

During compilation, \FS propagates individual Pauli errors through the Clifford circuit and records which measurement outcomes they flip.
The binary matrix $D$ has one row per detector, with a 1 for each measurement included in that detector's parity.
Each column of the binary matrix $\Omega$ marks the measurements flipped by one elementary error.
The detector signature matrix is
\begin{equation}
  H=D\Omega \pmod 2,
  \label{eq:dem-matrix}
\end{equation}
where each column marks the detectors flipped by the corresponding error.
A detector flips exactly when an error flips an odd number of the measurements in its parity.
The same construction using logical observable parities gives the logical signatures.

Channels with mutually exclusive Pauli errors must be expressed as independent Bernoulli events for this sampling procedure.
\FS performs this conversion for uniform depolarizing channels and supports one-qubit Pauli mixtures only when an exact representation exists.

At run time, \FS samples an event mask for each independent DEM event and XORs it into the detector and observable masks specified by that event's signature.
Repeated sampling therefore requires only Bernoulli sampling and bitwise XOR operations; it does not require propagating Pauli errors through the Clifford circuit again.

\subsection{Decoding and sensitivity estimation}

Both execution paths pass their packed detector records to a decoder.
The decoder returns a correction bit $c_{\mathcal D}(\bm d_i)$ for each shot $i$, indicating whether its logical readout should be flipped.
These bits are stored in the same packed format as a \emph{correction mask}.
A bitwise XOR of the observable and correction masks gives the \emph{loss mask} $(L_1,\ldots,L_N)$: bit $i$ is 1 exactly when the corrected readout in shot $i$ still differs from its noiseless value (\cref{eq:loss}).
Counting its 1 bits and dividing by $N$ gives the logical error estimate in \cref{eq:logical-error-estimate}.

Counting the 1 bits in the event mask for event $j$ gives the number of shots in which that event occurs.
The bitwise AND of the event and loss masks marks the shots for which both $F_{ij}=1$ and $L_i=1$.
Counting the 1 bits in this result gives the number of logical failures among shots containing event $j$.
Subtracting these two counts from $N$ and from the total failure count, respectively, gives the number of shots without the event and the number of logical failures in those shots.

For shots with and without event $j$, the estimator first subtracts $b$ times the shot count from the corresponding logical failure count.
It then weights the two adjusted counts by $1/p_j$ and $-1/(1-p_j)$, respectively, and divides their sum by $N$ to obtain the sensitivity estimate for event $j$ in \cref{eq:mc-estimator}.

\subsection{Computational cost and performance optimizations}
\label{sec:performance-mechanisms}

The score estimator in \cref{eq:mc-estimator} computes all $M$ sensitivities from one batch of $N$ sampled and decoded shots.
We denote the total runtime after compilation by $T_{\mathrm{score}}$.

Let $T_{\mathrm{sample}}(N)$ be the time to generate the simulation records for $N$ shots, and $T_{\mathrm{decode}}(N)$ the time to decode those records.
With $B=\lceil N/64\rceil$ 64-bit words per shot mask, a simplified cost model is
\begin{equation}
  T_{\mathrm{score}}
  =T_{\mathrm{sample}}(N)+T_{\mathrm{decode}}(N)+O(MB).
  \label{eq:attribution-cost}
\end{equation}
The $O(MB)$ term accounts for scanning the $M$ event masks, each containing $B$ words, together with the loss mask to compute the sensitivities.

To reduce the computational cost of this workflow, \FS combines circuit compilation with bit-parallel execution at run time.
During compilation, it derives the ideal measurement relations for the forward path and the error signatures for the DEM path, avoiding repeated circuit analysis across shots.
At run time, packed masks allow bitwise operations to update multiple shots at once, while population counts provide the statistics needed for sensitivity estimation.

The implementation further reduces overhead from memory allocation,
data conversion, and repeated setup.
Dense integer identifiers, preallocated buffers, and inline storage
for short symbolic expressions limit allocation and lookup costs.
DEM sampling can produce the decoder's requested data layout directly,
avoiding unnecessary conversions.
Collection workflows reuse compiled samplers across decoder tasks
and decoder workers across batches.
\Cref{tab:performance-mechanisms} summarizes the main performance
optimizations in \FS.

\section{Performance benchmarks}
\label{sec:benchmark}

We first compare the sampling performance of \FS and other stabilizer simulators on random Clifford circuits and surface code memories.
For a fair comparison, all simulators generate measurement records from the same input circuits, and \FS uses its forward stabilizer simulation path without recording error events or computing sensitivities.
We then use the same forward path to measure the additional runtime required for error attribution.
All benchmarks use CPU backends; \cref{app:benchmark-environment} summarizes the hardware, software versions, and execution settings.

\subsection{Random Clifford circuits}
\label{sec:random-clifford-benchmark}

We compare FaultScope, Stim, Qiskit Aer (CPU), Cirq, QuantumClifford, and SymFT (CPU) \cite{gidney2021stim,javadiabhari2024qiskit,cirq2025,quantumclifford2026,fang2026symft} on random Clifford circuits without stochastic errors.
The benchmark covers seven circuit sizes, ranging from 16 to 1024 qubits.
For each size, we generate three circuits, each consisting of 64 random layers.
In each random layer, we independently choose $H$, $S$, $S^\dagger$, or identity for each qubit, then randomly pair the qubits and choose CNOT, CZ, SWAP, or identity for each pair.
All simulators receive the same circuit instances, expressed using the common gate set $\{H,S,\mathrm{CNOT}\}$.
Each circuit ends with one $Z$ measurement per qubit in a fixed order, and each simulator samples 1,000,000 shots per instance.

We measure the total runtime from reading the circuit description from disk to completing the processing of the sampled measurement outcomes.
The timed interval includes input parsing, circuit construction in each simulator's representation, any required transpilation or sampler compilation, generation of 1,000,000 shots, and conversion and aggregation of the measurement records.
Thus, circuit preparation and output processing are included in the reported runtime.
Process startup, package imports, installation and compilation of the simulator software itself, and warm-up are excluded.

We first calculate the median runtime for each circuit instance.
For each simulator and circuit size, we report the median of these three values.
A size is included in the runtime comparison only if all three instances complete successfully.

\Cref{fig:stabilizer-benchmark-scaling}(a) shows how runtime scales with circuit size, and \cref{tab:stabilizer-benchmark-times} lists representative timings.
FaultScope, Stim, QuantumClifford, and SymFT complete all seven sizes, while Qiskit Aer completes five and Cirq two.
FaultScope has the lowest observed median runtime at every size.
For the largest circuits, with $n=1024$, the reported runtimes are 2.931~s for FaultScope and 5.182~s for Stim.
Across the three matched circuit instances, the speedup over Stim ranges from $1.61\times$ to $1.80\times$, with a median of $1.70\times$.

\subsection{Surface code memories}
\label{sec:surface code-benchmark}

We next compare the same simulators on rotated surface code memory-$Z$ circuits~\cite{dennis2002topological,tomita2014low} with independent circuit level Pauli noise at $p=10^{-3}$, using 1,000,000 shots per circuit.
The benchmark covers 12 code distances, ranging from 2 to 100.
At each distance, all six simulators receive the same circuit.
Each circuit has $d$ rounds, where $d$ is the code distance, and contains
\begin{equation}
  n_{\mathrm q}=2d^2-1,\quad
  n_{\mathrm m}=d^3+d^2-d,\quad
  n_{\mathrm{CX}}=4d^2(d-1)
  \label{eq:benchmark-sizes}
\end{equation}
qubits, measurements per shot, and CNOT operations, respectively.
At $d=100$, these counts are 19,999 qubits, 1,009,900 measurements per shot, and 3,960,000 CNOTs.

Single- and two-qubit gates are followed by depolarizing errors with total nonidentity probability $p$, distributed uniformly over the three or fifteen nonidentity Pauli operators, respectively.
Data qubits additionally undergo single-qubit depolarization before each syndrome extraction round; preparation errors and measurement bit flips each occur with probability $p$.
All six simulators implement the same noise channels.
We measure runtime from loading the circuit to completing the processing of its sampled measurement outcomes.
For each simulator and code distance, we report the median runtime and compare simulators at the same distance.

\begin{figure*}[t]
  \centering
  \includegraphics[width=0.99\textwidth]{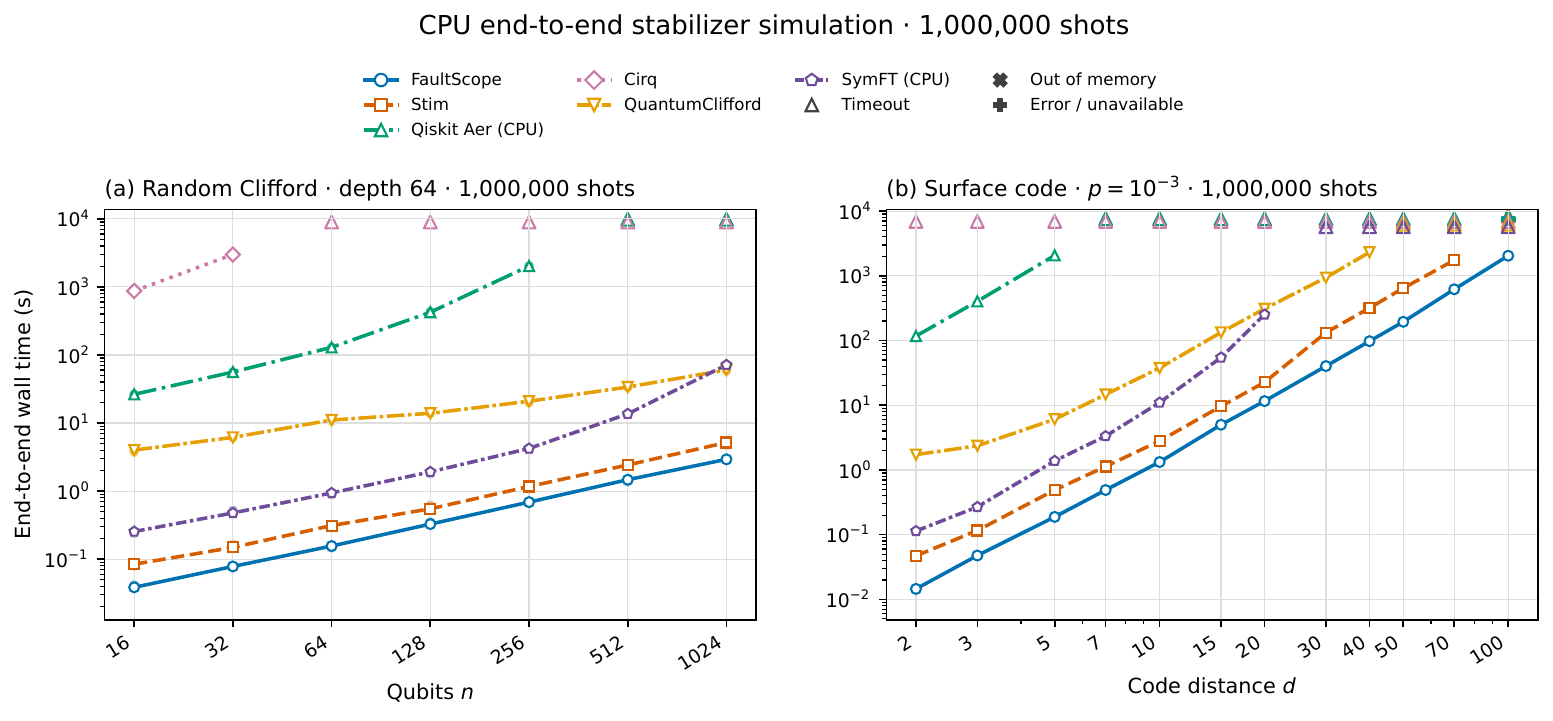}
  \caption{End to end wall time for CPU simulation of two workloads, using 1,000,000 shots per circuit.
  All simulators use batches of at most 1,000 shots and the same per-worker resource limits.
  (a) Random Clifford circuits with 64 random layers and no stochastic errors.
  (b) Rotated surface code memory-$Z$ circuits with $d$ rounds and circuit level Pauli noise at $p=10^{-3}$.
  Points show median runtimes.
  In both panels, open triangles at the top mark 3600 second timeouts,
  and filled plus signs mark other failures.}
  \label{fig:stabilizer-benchmark-scaling}
\end{figure*}

\Cref{fig:stabilizer-benchmark-scaling}(b) shows the scaling with code distance, with representative timings in \cref{tab:stabilizer-benchmark-times}.
FaultScope completes all 12 distances, Stim completes 11, QuantumClifford nine, SymFT seven, and Qiskit Aer three; Cirq times out at every distance.
FaultScope has the lowest observed median runtime at all 11 distances with a completed comparison, with speedups over Stim ranging from $1.93\times$ to $3.31\times$.
At $d=20$, the noisy circuit median runtimes are 11.59~s for FaultScope and 22.62~s for Stim, corresponding to a speedup of $1.95\times$.
SymFT takes 252.5~s at $d=20$ and times out at all five tested distances $d\geq30$.
At $d=70$, FaultScope and Stim take 618.8~s and 1741.5~s, respectively.
FaultScope is the only simulator to complete the $d=100$ circuit, taking 2043.1~s.
Timeouts and errors are excluded from the numerical runtime comparisons.

\begin{table*}[t]
\centering
\caption{Median end to end wall time (seconds) for CPU simulation at selected circuit sizes, using 1,000,000 shots per circuit in batches of at most 1,000 shots.
Timings include circuit preparation and compilation, sampling, and output processing.
TO denotes a 3600 second timeout; ERR denotes an execution error.
A $\dagger$ marks an entry for which at least one circuit instance has only one timing measurement.}
\label{tab:stabilizer-benchmark-times}
\begin{tabular}{lrrrr}
\toprule
 & \multicolumn{2}{c}{Random depth 64, $p=0$} & \multicolumn{2}{c}{Surface code, $p=10^{-3}$} \\
\cmidrule(lr){2-3}\cmidrule(lr){4-5}
Simulator & $n=128$ & $n=1024$ & $d=20$ & $d=100$ \\
\midrule
FaultScope & 0.3277 & 2.931 & 11.59 & 2043.1\textsuperscript{$\dagger$} \\
Stim & 0.5501 & 5.182 & 22.62 & TO \\
Qiskit Aer (CPU) & 425.0\textsuperscript{$\dagger$} & TO & TO & ERR \\
Cirq & TO & TO & TO & TO \\
QuantumClifford & 13.92 & 60.21 & 309.7\textsuperscript{$\dagger$} & TO \\
SymFT (CPU) & 1.918 & 71.49 & 252.5 & TO \\
\bottomrule
\end{tabular}
\end{table*}


\subsection{Additional cost of error attribution}
\label{sec:attribution-runtime}

To quantify the additional runtime required for error attribution, we compare the full sensitivity workflow with logical error rate estimation alone.
We use repetition code memories with $d=7,31,127,511,2047$ data qubits, $M=d$ error locations, and $N=100{,}000$ shots per evaluation.
Each data qubit is prepared in $\lvert0\rangle$.
Immediately after preparation and before any parity check gates, an $X$ error is independently applied to each data qubit with probability $p=0.5$; all other operations are ideal.
For each adjacent pair of data qubits, we measure the parity $Z_jZ_{j+1}$ using an ancilla qubit.
All data qubits are then measured individually in the $Z$ basis.
We use a fixed minimum weight decoder.
For each shot, it selects the $X$-error pattern that reproduces the measured parity check outcomes while involving the fewest data qubit errors.

The logical error rate only workflow disables event recording and returns the decoded failure frequency.
An auxiliary workflow retains events but does not compute sensitivities.
The attribution workflow uses the same circuit and decoder, retains error event records, and additionally computes all $M$ sensitivities with $b=0$.
For all three workflows, the measured runtime covers circuit and decoder setup, circuit compilation, sampling, decoding, and computation of the requested estimates.

\Cref{tab:attribution-runtime} reports total wall times, including construction and compilation for every evaluation.
At $M=2047$, the attribution workflow takes $4.074$~s, compared with $4.053$~s for the logical error rate only workflow.
Across the five sizes, computing the logical error rate and all $M$ sensitivities takes 0.33--0.97\% longer than computing the logical error rate alone without recording error events.

\begin{table}[!htbp]
  \centering
  \small
  \setlength{\tabcolsep}{4pt}
  \caption{Median total wall time in milliseconds for the repetition code experiment.
  Every evaluation uses $100{,}000$ shots.
  LER denotes the logical error rate; its baseline does not retain error events.
  LER+events retains error events without computing sensitivities.
  The attribution workflow also computes every sensitivity.
  All three use the same sampling API and packed output format.
  $\Delta$ is the attribution versus LER percentage difference calculated from the unrounded medians.
  Times include construction and compilation and are medians of seven repetitions.}
  \label{tab:attribution-runtime}
  \begin{tabular}{@{}rrrrr@{}}
    \toprule
    $d=M$ & LER & LER+events & Attribution & $\Delta$ (\%) \\
    \midrule
    7    & 7.205 & 7.241 & 7.274 & $+0.97\%$ \\
    31   & 32.000 & 32.080 & 32.249 & $+0.78\%$ \\
    127  & 130.891 & 130.922 & 131.320 & $+0.33\%$ \\
    511  & 561.740 & 563.784 & 565.945 & $+0.75\%$ \\
    2047 & 4052.660 & 4062.103 & 4073.938 & $+0.53\%$ \\
    \bottomrule
  \end{tabular}
\end{table}

\section{Discussion}
\label{sec:discussion}

\FS allows a QEC performance evaluation to guide subsequent noise reduction.
Component sensitivities connect the estimated logical error rate to the noisy operations that can be targeted for improvement.
\FS estimates all sensitivities from the same sampled and decoded shots, without separate circuit evaluations or changes to the decoder.
The benchmarks show fast simulation on random Clifford and surface code circuits, and low attribution overhead in the tested repetition codes.

At low logical error rates, a fixed shot budget may be insufficient to resolve the sensitivity differences needed to select between interventions.
A direction for future work is to adapt rare event sampling methods~\cite{bravyi2013rare} to sensitivity estimation, with the aim of improving precision when logical failures are scarce.

\section*{Data and code availability}

\FS is released at \url{https://github.com/yuguoshao/FaultScope}.
The benchmark scripts used in this study will be included in the same repository.

\section*{Use of generative AI}

Generative AI was used to assist with manuscript writing and code development.

\acknowledgments{
YS, NW, SC, ZW, and ZL are supported by the Beijing Natural Science Foundation (Grant No. Z220002).
YS is also supported by a startup funding from YMSC, Tsinghua Shuimu Scholar Program, and China Postdoctoral Science Foundation (Grant No.~2026M793709).
SC is also supported by the National Natural Science Foundation of China (Grants No.~12004205 and No.~12574253).
ZW is also supported by National Natural Science Foundation of China (Grant Nos. 62272259 and 62332009), and Beijing Science and Technology Planning Project (Grant No. Z25110100810000).
ZWL is supported in part by National Natural Science Foundation of China (Grant No.~12475023), Dushi Program, and a startup funding from YMSC.
}
\bibliographystyle{quantum}
\bibliography{main}

\clearpage
\onecolumn
\appendix

\section{Minimal forward attribution example}
\label{app:api}

The following example uses the public API to estimate sensitivity to one Bernoulli Pauli error, as defined in \cref{sec:attribution}.
For a qubit initialized in $\lvert0\rangle$, an $X$ error before a $Z$ measurement flips the measured bit.
Using that bit as the loss gives $L=F$, $\PL=p=0.2$, and $S=1$.
\begin{lstlisting}[language=Python]
from faultscope import (
    BernoulliPauliNoise, Circuit, FaultScopeSimulator,
    NoiseLocation, Operation,
)
noise = NoiseLocation(
    "data_x0", BernoulliPauliNoise("X"), 0.2, (0,)
)
circuit = Circuit(
    1, (
        Operation.noise(noise),
        Operation.measure(0, key="m0", basis="Z"),
    ),
)
result = FaultScopeSimulator(circuit).estimate(
    shots=1_000_000,
    seed=13,
    loss_mask_fn=lambda batch: batch.measurements["m0"],
    baseline="mean",
    top_k=1,
)
print(result.mean_loss)
print(result.sensitivities["data_x0"])
\end{lstlisting}

\section{Benchmark environment and artifact provenance}
\label{app:benchmark-environment}

\Cref{tab:stabilizer-benchmark-environment} summarizes the hardware and execution environments for all experiments.
The random Clifford and noisy surface code benchmarks run on the same host, with all six simulators included in each workload's comparison batch.
The two workload batches run sequentially with the same simulator installations and per-worker resource limits.
The surface code adapters are extended to support stochastic Pauli errors.
The repetition code experiments also use this host and the same FaultScope build.
\begin{table}[!ht]
\centering
\caption{Hardware, software environments, and execution settings.}
\label{tab:stabilizer-benchmark-environment}
\begin{tabular}{@{}p{0.29\linewidth}p{0.66\linewidth}@{}}
\toprule
{\raggedright Item\par} & {\raggedright Value\par} \\
\midrule
{\raggedright Processor (all experiments)\par} & {\raggedright Intel Xeon 6986P-C; 24 physical cores and 48 logical CPUs exposed to the host\par} \\
{\raggedright Host environment\par} & {\raggedright Debian~12; kernel \code{6.1.0-18-amd64}; Python~3.12.14\par} \\
{\raggedright Attribution execution\par} & {\raggedright Sequential evaluations; thread count environment variables set to one\par} \\
{\raggedright Suite CPU allocation\par} & {\raggedright Four physical cores (CPU IDs 0, 2, 4, 6)\par} \\
{\raggedright Surface code comparison\par} & {\raggedright Six simulators; 1,000,000 shots; up to four concurrent configurations\par} \\
{\raggedright Random Clifford comparison\par} & {\raggedright Six simulators; 1,000,000 shots; depth 64; up to four concurrent configurations\par} \\
{\raggedright Stabilizer controls\par} & {\raggedright CPU execution; one thread per worker; batches of at most 1000 shots; 36-GiB process-tree RSS threshold sampled every 50~ms; 3600-second timeout\par} \\
{\raggedright Other simulator releases\par} & {\raggedright Stim 1.16.0; Qiskit 2.5.0; Aer 0.17.2; Cirq 1.6.1; QuantumClifford 0.11.5; Julia 1.12.6; SymFT 0.1.1\par} \\
{\raggedright FaultScope (all experiments)\par} & {\raggedright 0.2.11\par} \\
{\raggedright Stim sampling path\par} & {\raggedright Neutral IR $\to$ Stim text $\to$ compile\_sampler $\to$ sample\_bit\_packed\par} \\
{\raggedright SymFT sampling path\par} & {\raggedright CPU batch sampler with packed measurement output; internal batch and chunk sizes capped at 1000 shots\par} \\
\bottomrule
\end{tabular}
\end{table}

The random Clifford grid contains 126 configurations: 105 complete and 21 time out.
The noisy surface code grid contains 72 configurations: 42 complete, 29 time out, and one returns an error when Qiskit Aer rejects the 19,999-qubit $d=100$ circuit against its 10,000-qubit target limit.
Neither grid has an out-of-memory or unsupported configuration.
For a given simulator, a random Clifford width is censored if any of its three instances fails to complete.

The 36-GiB memory threshold is applied to process-tree RSS samples every 50~ms rather than by a strict OS allocation cap.

For the noisy comparison, FaultScope and Stim use their built-in Pauli channels, while the Aer adapter converts the total nonidentity probability to Aer's depolarizing parameter convention.
The Cirq adapter samples two-qubit Pauli errors through its gate action interface, and the QuantumClifford adapter supplies the corresponding noise operation for Pauli frame sampling.
QuantumClifford evaluates one noiseless reference trajectory per timed invocation and reuses it across all shot batches within that invocation; the reference cost is included in the runtime.
The reported runtimes include these adapter costs.

\end{document}